\documentstyle[a4]{article}
\def\maketitle
{
\begin{flushright}
OUCMT-95-6
\end{flushright}
\begin{center}{\Large
\title
}
\end{center}
\begin{center}
\author\\
\vspace*{0.5cm}
{\it\inst}
\end{center}
\begin{quotation}
\abst\\
KEY WORDS: {\it\kword}
\end{quotation}
}

\def\citen#1{\cite{#1}}

\def\title{
Density Fluctuations in Traffic Flow
}

\def\author{
Satoshi {\sc Yukawa}
\footnote{e-mail:yuk@phys.sci.osaka-u.ac.jp}
and Macoto {\sc Kikuchi}
\footnote{e-mail:kikuchi@phys.sci.osaka-u.ac.jp}
}

\def\inst{
Department of Physics, Graduate School of Science,
Osaka University, \\Toyonaka 560, Japan
}

\def\abst{ Density fluctuations in traffic current are studied  
by
computer simulations using the deterministic coupled map
lattice model on a closed single-lane circuit.  By
calculating a power spectral density of  temporal density
fluctuations at a local section, we find a power-law
behavior, $\sim 1/f^{1.8}$, on the frequency $f$, in
non-congested flow phase.  The distribution of the headway
distance $h$ also shows the power law like $\sim 1/h^{3.0} $
at the same time.  The power law fluctuations are destroyed by
the occurence of the traffic jam.  }

\def\kword{
Traffic Flow, Density Fluctuation, 1/f noise, Coupled map  
lattice
}

\begin{document}
\sloppy
\maketitle

The traffic flow presents many interesting phenomena from the
viewpoint of physics: one of such phenomena is the $1/f$
fluctuation in the traffic current\cite{MH76,MH78}, which Musha
and Higuchi reported for the first time in 1976. They
recorded the transit time of the cars on a highway, and
calculated its power spectral density (PSD).  As a result,
they found a power law behavior like $ 1/f^\alpha$ on the
frequency $f$ in low frequency fluctuations and the white
noise behavior in high frequency fluctuations.  They got the
power $\alpha = 1 $ by fitting the data\cite{MH76} and got
$\alpha = 1.4$ by an analysis based on the Burgers
equation\cite{MH78}.

Recently, some attempts was made for understanding these
power-law fluctuations in the PSD of the traffic current
using computer simulations. Takayasu and Takayasu found
$1/f$ fluctuations in the jam phase of their cellular
automaton (CA) model by taking into account of the effects
of random braking\cite{TT93}. In the free flow phase, on the
other hand, no power law fluctuation was found.  In contrast
to their results, Nagel and Paczuski reported the $1/f$
fluctuations near the jamming transition point, using their
stochastic CA model\cite{NP95}.  These two studies are based
on stochastic CA models on the closed road. Quite recently,
another model which describes the two-lane traffic flow was
constructed based on a discrete-time continuous-space CA
(that is, the coupled map model, in our terminology) by
Zhang and Hu\cite{ZH95}; they also found the $1/f^{\alpha}$
type fluctuations in case of open road with the cars
injected from one end.

In this letter, we study the density fluctuations in a
traffic flow using the coupled map lattice (CML) model which
we proposed recently\cite{YK95}. In contrast to the CA
traffic flow models described above, the present model
follows the deterministic dynamics and is simulated on a
single-lane circuit.

Here we describe only the essence of the model; details are
found in the previous paper\cite{YK95}.  The model is based
on the coupled map lattice idea,\cite{Ka93} and thus is
constructed in the continuous space and the discrete time. A
continuous-space discrete-time model was also introduced by
Nagel and Herrmann as the infinite-state version of their
 CA model.\cite{NH93}. In the present model, each
vehicle is assigned a map which determines the velocity at
the next time step using the present values of the headway
distance and the velocity as inputs (We define the velocity
as the traveling distance during the unit time).  We put the
map describing a motion of a free vehicle in the following
form:
\begin{eqnarray}
v^{t+1} &=& F(v^t;v^F) \nonumber \\ & \equiv & \gamma
v^t + \beta \tanh \left ( \displaystyle
\frac{v^F-v^t}{\delta } \right) +\epsilon
\label{eq:freemap}
\end{eqnarray}
where $v^t $ is the velocity at time t.  $ v^F $ is the
preferred velocity, with which the vehicle tends to run when
no other car is found near ahead.
$\beta$,$\gamma$,$\delta$, and $\epsilon$ are controllable
parameters. The free map Eq.~\ref{eq:freemap} describes both
(1)the acceleration and the deceleration processes when
$v^t$ is away from $v^F$, and (2)the velocity fluctuation
around the preferred velocity when $v^t$ is close to $v^F$.
Throughout this paper, we take $\beta = 0.6$, $\gamma =
1.001$, $\delta= 0.1$, and $\epsilon=0.1$.  With these
parameters, the velocity fluctuation around $v^F$ is
expressed by a deterministic chaos.

We define interactions between the vehicles by introducing
two other maps, which are used according to the headway
distance.  Both maps determine the next value of the
velocity from the present values of the headway distance and
the velocity.  Suppose the vehicle A has the velocity
$v^t_A$ and is located at the position $x_A$ at $t$-th step,
and another vehicle B exists ahead of it at the position
$x_B ( > x_A ) $.  We define the headway distance $h$
as
\begin{eqnarray}
h = x_B-x_A+1,
\end{eqnarray}
where we take the length of the vehicles as the unit of
length. If $h < v^t_A $, the headway distance is too
small to be traveled with $v^t_A$ so that the vehicle A is
forced to apply sudden braking for avoiding a collision.
For expressing the sudden braking process, we employ the
similar procedure as what used in ref.~\citen{NS92}; that
is, the vehicle A changes its next velocity as $v_A^{t+1} =
h $.  We call this map the sudden braking map.  For $
v_A^t \le h < \alpha v_A^t~~~ (\alpha >1)$, the
vehicle changes its next velocity according to the following
equation:
\begin{eqnarray}
v^{t+1}_A &=& G(h, v^t_A;
v^F_A) \nonumber \\ & \equiv & \displaystyle
\frac{F(v^t_A;v^F_A) - v^t_A}{(\alpha -1) v^t_A}
(h -v^t_A) +v^t_A.
\end{eqnarray}
We call it the slowing-down map, which connects the free map
and the sudden braking map continuously.  When $h >
\alpha v_A^t$, the vehicle A can drive freely and thus the
free map Eq.~\ref{eq:freemap} is used.  We carry out the
simulation using these three maps taking $\alpha = 4.0$.

The simulations are made for a single-lane circuit with the
number of vehicles fixed. No overtaking is allowed
throughout.  As the initial state, the positions of the
vehicles are randomly chosen and the preferred velocity is
uniformly distributed in the range $[2.0,4.0]$. One step of
the simulation consists of the following three procedures:
first, the headway distance is measured for all vehicles.
Next, all the vehicles move simultaneously according to the
present velocity.  Finally the next velocity is determined
from the measured headway distance and the present velocity
using the maps introduced above.

We show results of the simulations.  The relation of the
concentration $\rho$ and the flow $q$ for the whole system,
which is called the q-k diagram or the fundamental diagram
in the traffic engineering, is shown in Fig.~1.
This diagram represents the macroscopic properties of the
traffic flow.
There is a concentration $\rho_c$ at which the flow $q$ takes
its maximum value\cite{YK95}. This concentration separates
two phases: the free flow phase in lower concentration $\rho
< \rho_c$, and the congested flow phase in higher
concentration $\rho > \rho_c$.  As we have shown in the
previous paper\cite{YK95}, a finite spatial region (or
regions, in cases of very high concentration) of the traffic
jam is formed in the congested flow phase. So we call
$\rho_c$ as the transition point of the jamming transition.

In order to study the density fluctuations, we make the time
series of the density measured at a local section in the
closed circuit. We take the length of the circuit as
$L=100000$ and the length of the local section for the
observations as 20. The PSD of the time series is calculated
by the Fourier transformations.  The result for $ \rho =
0.19 $, which is in the free flow phase, is shown in
Fig.~2.
Clear power law behavior like $1/f^{\alpha} $ on the frequency
$f$ is seen in the PSD.
We find that $\alpha \sim 1.8$.

The PSD in the congested flow phase for $\rho = 0.20 $ is
shown in Fig.~3.
In contrast with the free flow phase, no power-law behavior
is seen; white-noise-like behavior is observed instead,
within a wide range of the frequency.  To see more clearly
the difference in the PSD in these
two phases, we show two PSD's, both for the concentrations
much closer to $\rho_c$, in Fig.~4: one for
the concentration just below the transition, $\rho =0.197575
$, and another for the concentration just above the
transition, $ \rho =0.197576 $.
The figure shows that the power-law behavior is really seen
only below the transition and disappears above the
transition; in other words, long-time correlations in the
traffic current fluctuation persist in the free flow phase
and are destroyed by the occurence of the jamming
transition.

To study the origin of the power-low fluctuation in the free
flow phase, we make a snapshot of the density profile for $
\rho = 0.19 $ and $L=100000$, which is shown in
Fig.~5.  The density profile is calculated
by dividing the system into 1000 segments.
Thus the flucutuations in the scale shorter than 100 are
averaged out.
In this scale, a number of large clusters of the vehicles are
seen in the density profile.  Clearly, each
cluster is lead by the slowest vehicle within it,
that is, one having the slowest preferred velocity.
Other faster vehicles are forced to be stuck behind it.
Some clusters show internal density fluctuations
and some are non-fluctuating.
The average density in the non-fluctuating clusters is
found as $\sim 1/3 $, that is, the average
headway distance is $\sim 2$ there (the average
head-to-head distance is $\sim 3$).
The existence of this
characteristic headway distance is the origin of
the sharp high frequency peak seen in the PDS
in Fig.~2.  The low
frequency cutoff in the PDS, on the other hand,
is understood as the inverse of the average time which a
vehicle take to travel around the whole system.

We also show the histogram of the measured headway
distance $h$ in Fig.~6.
The characteristic headway distance 2 is seen
also in this figure.  Moreover, the headway
distance distribution shows a clear power-law
behavior of the form $1/h^{\beta}$
with $\beta \sim 3.0$
in the range from $ h \sim
3 $ to $ h \sim 70 $, which corresponds to the
scale {\em within} the cluster in
Fig.~5.
Larger headway distances, on the other hand, correspond to the
inter-cluster scale.

It is clear that in the free flow phase of the present model
all the vehicles would eventually form a single large cluster,
which is headed by the slowest vehicle among all.
We, however, see a number of clusters
in Fig.~5.
Thus the whole system
has not yet reached the final stationary state. That
is, the power-law fluctuations found above are in
non-stationary states.
In fact, it is expected to take a very long time
for the stationary state to be achieved, since
there are a number of slow vehicles whose
preferred velocities are close to each
other.  At the same time, we have no reason
as well to expect that the stationary state is
achieved in the real traffic flow.
Therefore, we think that
the power-law fluctuations found in the
simulations are actually related to the
one observed in the real traffic flow.

To summarize, we found $1/f^{\alpha}$ type behavior in the
PSD of the temporal density fluctuations in the free flow
phase of the CML model.  The model follows a deterministic
dynamics; that is, it develops deterministically into a
state which shows the power-law fluctuation. At the same
time, the vehicles form the clusters, within which the
headway distance distribute with the power-law, $ \sim
1/h^{3.0} $. This power law distribution is considered as
the origin of the $1/f^{\alpha}$ fluctuation.  In the
congested flow phase, on the other hand, the power law
fluctuation is not seen; in other words, the occurence of
the traffic jam destroys long time correlations in the
traffic current fluctuation.  It is in contrast with the
result by Takayasu and Takayasu,\cite{TT93} in which the
power law was observed only in the congested phase, and also
with the one by Nagel and Paczuski\cite{NP95}, who found the
power law only near the transition point. The reason why the
long-time correlations are destroyed by the traffic jam in
the present model is understood as follows: Once a vehicle
runs into a jam region, it is forced to stop.  When running
out of the jam region, it starts again from the beginning so
that the memory of the current fluctuations is lost then.
We obtained $ \alpha \sim 1.8 $, the power which is somewhat
larger than the values obtained from the real traffic flow,
$ 1$ or $1.4 $, or ones observed by other
simulations\cite{TT93,NP95,ZH95}. Origin of the difference in  
the values
of $\alpha$ is not yet clear.  Other sort of power-law
phenomena than the $1/f^{\alpha} $ current fluctuation were
also reported by some simulations\cite{NH93,Le94,NP95}. To  
clarify their
relations is also a problem to be left for future studies.

A subject related to the density fluctuations
in the traffic flow is the $ 1/f^{\alpha} $
fluctuations in granular flows:
By experiments of
a narrow channel like a pipe or a hourglass, the
power-law fluctuations like $ \alpha = 1 $ or $
\alpha =1.5 $ were
observed\cite{SV74,HNNM95}.
Computer simulations were made recently to reproduce these
power law.\cite{PH94,Po94,RH94,PH95}.
It would be worth pointed out that the real experiments
of granular flows
are made in non-stationary states of the flow,
as in the case of the present study.
The traffic flow and the granular flow share several
similar properties, and have many different
points as well.
To make their relations clear is a
challenging problem.  Especially, we found the
power-law distribution of the headway distance,
which we think is the origin of the
$1/f^{\alpha}$ fluctuation.  It would be
interesting to study corresponding quantity in
the granular flow.

We are grateful to Y. Akutsu, K. Tokita, and
T. Nagao for valuable discussions. We also thank
S. Tadaki, M. Bando, K. Hasebe, M.  Kawahara, H.
Ozaki, N. Ito, M. Takayasu and H. Takayasu for
valuable discussions. We are grateful to
H. Herrmann for sending us of ref.~\citen{NH93}.
Numerical calculation were done mainly by
Fujitsu VPP 500 in the Institute for Solid State
Physics, the University of Tokyo and NEC
SX-3/14R in the computer center of Osaka
University.

\leftline{\Large Figure Captions}

\begin{description}
\item[Fig. 1]
Plot of the flow $q$ against the concentration
$\rho$ measured for the whole
system. The length of the system is 1000.
Average over 1000 steps are taken
after 5000 steps being discarded.
The average over 10 samples are shown.

\item[Fig. 2]
The log-log plot of the power spectral
density of the temporal density fluctuation.
Concentration of vehicle is $ \rho = 0.19 $. The system
length $L $ and a length of observation section
are 100000 and 20, respectively.
We record a time series of 65536 steps after discarding 400000  
steps.
The average over 30 samples are shown.
The slope of the straight line is $ -1.8 $.

\item[Fig. 3]
The log-log plot of the power spectral
density of the temporal density fluctuation for $ \rho = 0.20 $.
The length parameters and the simulation steps are same as
Fig.~2.
The result of a single sample is shown.

\item[Fig. 4]
The log-log plot of the power spectral
densities. The system length is $L=200000$ and the length of
local section for observation is 20.
We record the time series for 32768 steps after discarding  
800000
steps.  No sample average is taken.
(a)$\rho =0.197575 $, which is just below the transition.
(b)$ \rho =0.197576 $, which is just above the transition.

\item[Fig. 5]
Snapshot of the density profile for $L= 100000$
and $ \rho = 0.19 $.
We take it after discarding 500000 time steps.
Density is calculated by dividing the system into 1000
segments.

\item[Fig. 6]
The log-log plot of the distribution histogram of
the headway distance for $L=100000$ and $ \rho =0.19 $.
Distribution is calculated in 100 steps
after discarding 500000 steps.
The slope of the straight line is $ -3.0 $.
\label{fig:dishead}
\end{description}

\end{document}